\documentclass[oneside,12pt]{article}
\usepackage{amsmath,amsgen,amstext,amsbsy,amsopn,amsthm, amssymb}
\usepackage{latexsym}

\newtheorem{thm}{Theorem}
\numberwithin{thm}{section}
\newtheorem{cor}{Corollary}
\numberwithin{cor}{section}
\newtheorem{lem}{Lemma}
\numberwithin{lem}{section}

\numberwithin{prop}{section}
\theoremstyle{definition}

\numberwithin{defn}{section}

\numberwithin{rem}{section}

\newcommand{\xij}{|x_i-x_j|}

\newcommand{\R}{\mathbb{R}}

\numberwithin{equation}{section}  
\def\mfr#1/#2{\hbox{$\frac{{#1} }{ {#2}}$}}

\begin{document}
\markboth{\scriptsize{LY 4/02/00}}{\scriptsize{LY 4/02/00}}
\title{\bf{The Ground State Energy of a Dilute Two-dimensional Bose Gas}}
\author{\vspace{5pt} Elliott H.~Lieb$^1$, and Jakob 
Yngvason$^{2}$\\ 
\vspace{-4pt}\small{$1.$ Departments of Physics and Mathematics, Jadwin 
Hall,} \\
\small{Princeton University, P.~O.~Box 708, Princeton, New Jersey
  08544}\\
\vspace{-4pt}\small{$2.$ Institut f\"ur Theoretische Physik, Universit\"at 
Wien}\\
\small{Boltzmanngasse 5, A 1090 Vienna, Austria}}
\date{February 4, 2000}
\maketitle

\footnotetext[1]{Work partially
supported by U.S. National Science Foundation
grant PHY 98-20650.\\
\copyright\,1999 by the authors. This paper may be reproduced, in its
entirety, for non-commercial purposes.}

\begin{abstract} The ground state energy per particle of a dilute,
homogeneous, two-dimensional  Bose gas, in the thermodynamic limit is
shown rigorously to be $E_0/N = (2\pi \hbar^2\rho /m){|\ln
(\rho a^2)|^{-1}}$, to leading order, with a relative error at most
${\rm O} \left(|\ln (\rho a^2)|^{-1/5}\right)$. Here
$N$ is the number of particles, $\rho =N/V$ is the particle density and
$a$ is the scattering length of the two-body potential. We assume that
the two-body potential is short range and nonnegative. The amusing
feature of this result is that, in contrast to the three-dimensional
case, the energy, $E_0$ is not simply $N(N-1)/2$ times the energy of two
particles in a large box of volume (area, really) $V$. It is much
larger.  \end{abstract}

\section{Introduction} An ancient problem, going back to the 1950's, is
the calculation of the ground state energy of a dilute Bose gas in   the
thermodynamic limit.  The particles are assumed to interact only with a
two-body potential and are enclosed in a box of side length $L$.  A
formula  was derived for the energy $E_0(N,L)$ in three dimensions for
a two-body potential $v$ with scattering length $a$ (see Appendix) and
fixed particle density $\rho = N/V$, ($N=$ particle number and $V=$
volume =$L^3$ in three dimensions). In the thermodynamic limit, the
energy/particle is

\begin{equation} e(\rho) \equiv \lim_{N\to \infty}E_0(N,
\rho^{-1/3}N^{1/3})/N
\simeq 4 \pi \mu \rho a
\label{3den}
\end{equation}
to lowest order in $\rho$. Here, $\mu= \hbar^2/2m$ with $m$ the
mass of a particle.

Our goal here is to derive the analogous low density formula for a
two-dimensional Bose gas.

There were several approaches in the 50's and 60's to the derivation of
the three-dimensional formula \eqref{3den}, but none of them were
rigorous.  Recently we were able to give a rigorous derivation of
(\ref{3den}) and we refer the reader to \cite{LY1998} for a physically
motivated discussion of the essential difficulty in proving
\eqref{3den}, which, basically, is the fact that at low density the
mean interparticle spacing is much smaller than  the mean de Broglie
wavelength of the particles. Thus, Bose particles cannot be thought of
as localized. Furthermore, in \cite{LY1998}, we explain rather
carefully why the usual expression `perturbation theory' is not
appropriate for (\ref{3den}) --- especially in the hard core case.
Indeed, Bogolubov's 1947 'perturbation theory' \cite{BO} yields an
estimate, which is incorrect for the low density limit:

\begin{equation} 
e(\rho) \simeq \frac{1}{2}\rho \int_{\R^{3}} v . 
 \end{equation} 

It was only with a leap of faith that Bogliubov and Landau recognized
that $\int v$ is the first Born approximation to $8\pi \mu a$ and thus
were able to derive \eqref{3den}. Obviously this cannot be called
perturbation theory.  Moreover, depending on the nature of $v$,  it is
sometimes the potential energy and sometimes the kinetic energy that is
the dominating quantity; for example, in the hard core case the kinetic
energy is  the perturbation, rather than the potential energy, as the
Bogolubov method assumes.

The two-dimensional theory, in contrast, began to receive attention
only much later. The first derivation of the correct asymptotic formula
was, to our knowledge, done by Schick \cite{schick} for a gas of hard 
discs:

\begin{equation}
    e(\rho)  \simeq 4\pi \mu \rho |\ln(\rho a^2) |^{-1}.
\label{2den}
\end{equation}

This was accomplished by an infinite summation of `perturbation series'
diagrams. Subsequently, a corrected modification of \cite{schick} was
given in \cite{hines}. Positive temperature extensions were given in
\cite{popov} and in \cite{fishho}. All this work involved an analysis in
momentum space --- as was the case for \eqref{3den}, with the exception
of a method due to one of us that works directly in configuration space
\cite{lieb63}.  Ovchinnikov \cite{Ovch} derived \eqref{2den} by using,
basically, the method in \cite{lieb63}. Again, these derivations require
several unproven assumptions and are not rigorous. 

One of the intriguing facts about \eqref{2den} is that the energy for $N$
particles is {\it not equal} to $N(N-1)/2$  times the energy for two
particles in the low density limit --- as is the case in
three dimensions.  The latter quantity,  $E_0(2,L)$, is, asymptotically
for large $L$, equal to $8\pi \mu L^{-2} \left[ \ln(L^2/a^2)
\right]^{-1}$.  Thus, if the  $N(N-1)/2$ rule were to apply, \eqref{2den}
would have to be replaced by the much smaller quantity $4\pi \mu
\rho\left[ \ln(L^2/a^2) \right]^{-1}$. In other words, $L$, which tends
to $\infty$ in the thermodynamic limit, has to be replaced by the mean
particle separation, $\rho^{-1/2}$ in the logarithmic factor. Various
poetic formulations of this curious fact have been given, but the fact
remains that the non-linearity is something that does not  occur in more
than two-dimensions and its precise nature is hardly obvious,
physically. This anomaly is the main reason that the present
investigation is not a trivial extension of \cite{LY1998}.

We will prove \eqref{2den} for nonnegative, finite range two-body
potentials by finding upper and lower bounds of the correct form.  The
restriction to finite range can be relaxed somewhat, as  was done in
\cite{LSY1999}, but the restriction to nonnegative $v$ cannot be
removed in the current state of our methodology.  The upper bounds will
have relative remainder terms O($|\ln(\rho a^2)|^{-1}$) while the lower
bound will have remainder O($|\ln(\rho a^2)|^{-1/5}$).  It is claimed
in \cite{hines} that the relative error for a hard core gas is negative
and O($\ln |\ln(\rho a^2)||\ln(\rho a^2)|^{-1})$, which is consistent
with our bounds.

In the next section we shall give the upper bound (following Dyson's
analysis  for the three-dimensional hard core gas \cite{dyson}). Then
we shall recall our method in \cite{LY1998} for the lower bound and
show how it has to be modified. An important point concerns the
definition of the scattering length in two dimensions (which will be
discussed in detail in Appendix A) and how `Dyson's Lemma' \cite{dyson,
LY1998} has to be  modified accordingly (Appendix B).

An obvious extension of the present work is the case of 2D bosons in a trap
and this will be the subject of a forthcoming paper. Just as the 
passage from 3D to 2D for the homogeneous case presents some non-trivial
issues that have to be resolved, so the correct generalization of the
Gross-Pitaevskii equation \cite{DGPS} to the 2D dilute trapped gas presents some
additional complications.

We thank P. Kevrekidis for drawing our attention to this problem.

\section{ Upper Bound for the Ground State Energy}
We begin with the well known definition of the Hamiltonian under
discussion:
\begin{equation}\label{ham}
H^{(N)}=-\mu\sum_{i=1}^{N}\nabla_i^2 +\sum_{i<j}v(\xij),
\end{equation}
We assume that $v(r)\geq 0$ and $v(r)=0$ if $r>R_0$, for some 
$R_0<\infty$.  The Hamiltonian \eqref{ham} acts on totally {\it 
symmetric}, square integrable wave functions of $(x_1,\dots,x_N)$ with 
$x_i\in\R^{2}$.  Its ground state energy in a box (rectangle, 
actually) of side length $L$ is
\begin{equation}
E_0(N,L)=\inf_{\Psi}\frac{\langle\Psi,H^{(N)}\Psi\rangle}
{\langle\Psi,\Psi\rangle}
\end{equation}
where the infimum is over all wave functions $\Psi$ satisfying
appropriate conditions on the boundary of the box. For the
upper bound it is natural to use Dirichlet boundary conditions, which gives 
the 
largest energy, but for the actual calculations it is more convenient
to use periodic boundary conditions and a periodic extension of the 
interaction potential. This can only raise the energy since $v\geq 0$.
 Localization of the wave functions on the
length scale $L$ to obtain Dirichlet boundary conditions costs an energy
$\sim{(\rm (const.)}\,L^{{-2}}$ per particle, so in the 
thermodynamic limit our upper bound is also  a valid upper 
bound  for Dirichlet 
boundary conditions.  For the lower bound, on the other hand, we shall 
use Neumann boundary conditions,  which yields  the smallest energy.

Following \cite{dyson} we make a variational  ansatz for $\Psi$ of 
the following form:
\begin{equation}
\label{ansatz}\Psi(x_{1},\dots,x_{N})=\prod_{i=2}^Nf(t_i(x_1,\dots,x_i))
\end{equation}
where $t_i = \min\{|x_i-x_j|, 1\leq j\leq i-1\}$ is the distance 
of $x_{i}$ to its {\it nearest neighbor} among the points 
$x_1,\dots,x_{i-1}$ and $f$ is a  nondecreasing function of 
$t\geq 0$ with values between zero and 1. 

We wish to calculate $\langle\Psi,H^{(N)}\Psi\rangle
/{\langle\Psi,\Psi\rangle}$.  Dyson \cite{dyson} carried out this 
calculation for the hard core case, namely when $f(r)=0$ for $r<$ the 
core radius.  His formula has been generalized in \cite{LSY1999} in 
two directions: One is the inclusion of an external potential (which 
we do not need here) and the other (which we do need) is the extension 
to a non-hard core potential $v$.  We refer to \cite{LSY1999} for 
details.  The result involves the following three integrals
\begin{eqnarray}I &=& 2\pi\int_{0}^{\infty}(1-f(r)^{2})r\,dr\\
J&=& 
2\pi\int_{0}^{\infty}\left(|f'(r)|^{2}+\mfr1/2v(r)|f(r)|^{2}\right)r\,dr\\
K&=&2\pi\int_{0}^{\infty}f(r)f'(r)r\,dr
\end{eqnarray}
In terms of these integrals the bound on the energy is
\begin{equation}\label{upperijk}
\langle\Psi,H^{(N)}\Psi\rangle/ {\langle\Psi,\Psi\rangle} \leq N\left( 
\frac{\rho J}{1-\rho I} +\frac23\frac{(\rho K)^{2}}{(1-\rho 
I)^{2}}\right).
\end{equation}
The form of this bound is the same as in \cite{dyson}. Compared to Eq.
(3.29) in \cite{LSY1999} 
there is a factor
$(1-\rho I)^{-1}$ in the first term in place of $(1-\rho I)^{-2}$.
This can be traced to the use of the Cauchy-Schwarz 
inequality in Eq. (3.19) in \cite{LSY1999} which is not necessary in the
case of the homogeneous system treated here.

The next step is to make a choice for $f$, and this will involve the 
scattering length $a$ and a variational parameter $b$.  First, we 
have to define the scattering length.

Consider the Schr\"odinger equation 
\begin{equation}
\label{scatt}-\mu\Delta \phi_{0}+\mfr1/2v \phi_{0}=0.
\end{equation}
We do {\it not} require $\phi_{0}$ to be bounded.  As shown in
Appendix A, up to an overall factor there is a unique, nonnegative, 
spherically symmetric $\phi_0(x)=f_{0}(|x|)$ that satisfies \eqref{scatt} 
provided 
the Schr\"odinger operator $-\mu \Delta + \frac{1}{2} v(r)$ in 
$L^{2}(\R^{2})$ has no bound states.  For $r> R_{0}$, $f_0$ 
necessarily has the form (since $\phi_{0}$ is a harmonic function outside 
the range of $v$)
\begin{equation}\label{scattl}
f_{0}(r)=\hbox{\rm (const.)\,}\ln(r/a).
\end{equation}
The length $a$ is called the {\it scattering length}.  Note that it 
depends on both $\mu$ and on $v$.  In the case that $v$ is 
nonnegative, $f_{0}$ is necessarily a monotonically increasing 
function of $r$.

We now define our variational $f$ to be
\begin{equation}\label{deff}
f(r)=\begin{cases}
f_0(r)/f_{0}(b)&\text{for $0\leq r\leq b$},\\
1&\textrm{for $r>b$}.
\end{cases}
\end{equation}
with some $b>R_{0}>a$ to be chosen in an optimal way. 
By Appendix A we have that
$f$ satisfies $f'\geq 0$ and $0\leq f\leq 1$, for all $b$.
Moreover, 
\begin{equation}
f(r)\geq \begin{cases}
\ln (r/a)/\ln (b/a)&\text{for $a\leq r\leq b$},\\
0&\textrm{for $r<a$}.
\end{cases}
\end{equation} 
Using this information one computes
\begin{eqnarray}I &\leq &\pi{a^2}+2\pi\int_{a}^b\left(1-
	\frac{[\ln (r/a)]^2}{[\ln (b/a)]^2}\right)rdr=
\frac{\pi b^{2}}{\ln(b/a)}\left(1+{\rm O}([\ln 
(b/a)]^{-1})\right)\nonumber\\ \\
J&=&2\pi \left[f(r)f'(r)r \right]_{0}^b =\frac{2\pi}{\ln(b/a)}
\\
K&=&\pi\int(f(r)^2)'r\,dr=\pi b-\pi\int f(r)^2dr\nonumber
\\ &\leq&\pi b-
\pi\int_{a}^b\frac{[\ln (r/a)]^2}{[\ln (b/a)]^2}dr
= \frac{2\pi b}{\ln(b/a)}\left(1+{\rm O}([\ln 
(b/a)]^{-1}\right).
\end{eqnarray}
Inserted in \eqref{upperijk} this leads to the upper bound 
\begin{equation}\label{upperbound}
E_{0}(N,L)/N\leq \frac{2\pi\rho}{\ln(b/a)-\pi\rho b^{2}}\left(1+{\rm 
O}([\ln(b/a)]^{{-1}})\right).
\end{equation}
The minimum over $b$ of the leading term is obtained for 
$b=(2\pi\rho)^{{-1/2}}$. Inserting this in \eqref{upperbound} we thus 
obtain
\begin{thm}[Upper bound]
The ground state energy with periodic boundary conditions satisfies
\begin{equation}
E_{0}(N,L)/N\leq \frac{4\pi\rho}{|\ln(\rho a^{2})|}\left(1+{\rm 
O}(|\ln(\rho a^{2})|^{{-1}}\right).
\end{equation}
\end{thm}
Dirichlet boundary conditions may introduce an additional relative
error, but as already noted it is at most  $\Delta E_0/N \propto
L^{-2}$.

\section{Lower Bound to the Ground State Energy}

The method of \cite{LY1998} for obtaining a lower bound to 
$E_{0}(N,L)$ involves the following steps:

\begin{itemize}
\item[{1} ]A generalization of a lemma due to Dyson \cite{dyson} that
allows the replacement of the interaction potential $v$ by a `soft'
potential $U$ at the cost of sacrificing kinetic energy.
\item[{2}] Division of the large box of side length $L$ into small
boxes of side length $\ell$, which is kept {\it fixed} as $L\to\infty$,
and a corresponding lowering of the energy by the use of {\it Neumann}
boundary conditions on each box.  It is necessary to minimize the total
energy over all distributions of the particles among the small boxes;
this is accomplished with the aid of the superadditivity of the ground
state energy in each box (i.e., $E_{0}(N_1+N_2,L) \geq
E_{0}(N_1,L)+E_{0}(N_2,L)$, which follows from $v\geq 0$).
\item[{3}] The use of a rigorous version of first order perturbation
theory, known as Temple's inequality \cite{TE}, to estimate from below
the energy with the new potential $U$ in the small boxes.
\end{itemize}

We follow the same strategy here,  but there are several 
modifications to be made.  The two dimensional version of the 
generalized Dyson Lemma is as follows.

\begin{lem}\label{dysonl} 
Let $v(r)\geq0$ and $v(r)=0$ for $r>R_0$.
 Let $U(r)\geq 0$ be any function satisfying
\begin{equation}\label{1dyson}
 \int_0^\infty U(r)\ln(r/a)rdr \leq 1~~~~~{\rm and}~~~~~ U(r)=0 ~~~{\rm
for}~
 r<R_0.
\end{equation}
Let ${\cal B}\subset \R^2$ be star-shaped  with respect
to $0$.
Then, for all functions $\phi  \in H^1(\mathcal{B})$,
\begin{equation}
\int_{\cal B} \mu|\nabla \phi (x)|^2 + \frac{1}{2}v(r)|\phi (x)|^2~d^2x  
\geq  \mu  \int_{\cal B} U(r)
|\phi (x)|^2 ~d^2x.
\label{2dyson}
\end{equation}
\end{lem}

A domain $\mathcal{B}$ is star-shaped with respect to a point $p$ if the
line segment $[p,x]\subset \mathcal{B}$ whenever $x \in \mathcal{B}$.  A
convex domain is star-shaped with respect to any point in it (and
conversely). The three-dimensional version of the lemma replaces 
\eqref{1dyson} with $\int_0^\infty U(r)r^2dr \leq a$.

\noindent The proof is given in Appendix B.

As in \cite{LY1998}, Lemma \ref{dysonl} can be used to bound the 
many body Hamiltonian $H^{{(N)}}$ from below, as follows:
\begin{cor}
For any $U$ as in Lemma \ref{dysonl} and any $0<\varepsilon<1$
\begin{equation}\label{epsilonbd}
H^{(N)} \geq \varepsilon T^{(N)} +(1-\varepsilon)\mu W
\end{equation}
with $T^{(N)}=-\mu\sum_{i=1}^{N}\Delta_{i}$ and 
\begin{equation}
\label{W}W(x_{1},\dots,x_{N})=\sum_{i=1}^{N}U\left(\min_{j,\,j\neq
i}|x_{i}-x_{j}|.\right).
\end{equation}
\end{cor} 

For $U$ we choose the following functions, parameterized by $R>R_{0}$:
\begin{equation}U_{R}(r)=\begin{cases}\nu(R)^{-1}&\text{for
$R_{0}<r<R$ }\\
0&\text{otherwise}
\end{cases}
\end{equation}
with $\nu(R)$ chosen so that
\begin{equation}
\int _{R_{0}}^{R}U_{R}(r)\ln(r/a)r\,dr=1
\end{equation}
for all $R>R_{0},$
i.e.,
\begin{equation}\label{nu}
\nu(R)=\int_{R_{0}}^{R}\ln(r/a)r\,dr=\mfr1/4 \left\{R^{2}
\left(\ln(R^{2}/a^{2})-1\right)-R_{0}^{2}
\left(\ln(R_{0}^{2}/a^{2})-1\right)\right\}.
\end{equation}
The nearest neighbor 
interaction \eqref{W} corresponding to $U_{R}$ will be denoted $W_{R}$.

As in \cite{LY1998} we shall need estimates  on the expectation
value, $\langle W_R\rangle_{0}$,  of $W_{R}$ in the ground state of
$\varepsilon T^{(N)}$ of \eqref{epsilonbd} with
Neumann boundary conditions. This is just the average value of $W_{R}$
in a hypercube in $\R^{2N}$.  Besides the
normalization factor $\nu(R)$, the computation involves 
the volume (area) of the support of
$U_{R}$, which is
 \begin{equation} A(R)=\pi(R^{2}-R_{0}^{2}).
\end{equation}

In contrast to the three-dimensional situation the normalization factor
$\nu(R)$ is not just a constant ($R$ independent) multiple of $A(R)$;
 the factor $\ln(r/a)$ in \eqref{1dyson} accounts for the more
complicated expressions in the two-dimensional case.   Taking into
account that $U_{R}$ is proportional to the characteristic function of 
a disc of radius $R$ with a hole of radius $R_{0}$, the following 
inequalities for $n$ particles in a box of side length 
$\ell$ are obtained by the same geometric reasoning as in 
\cite{LY1998}:
\begin{eqnarray}\label{firstorder}
\langle 
W_R\rangle_{0}&\geq&\frac {n}{\nu(R)}
\left(1-\mfr {2R}/{\ell}\right)^2\left[1-(1-Q)^{(n-1)}\right]\\
\langle 
W_R\rangle_{0}&\leq&
\frac 
{n}{\nu(R)}\left[1-(1-Q)^{(n-1)}\right]
\end{eqnarray}
with 
\begin{equation}
Q=A(R)/\ell^{2}
\end{equation}
  being   the relative volume occupied by the
support of the potential $U_{R}$.
Since $U_{R}^{2}=\nu(R)^{-1}U_{R}$ we also have
\begin{equation}
\langle 
W_R^{2}\rangle_{0}\leq \frac n{\nu(R)}\langle 
W_R\rangle_{0}.
\end{equation}

As in \cite{LY1998} we estimate $[1-(1-Q)^{(n-1)}]$ by
\begin{equation}
(n-1)Q\geq \left[1-(1-Q)^{(n-1)}\right]\geq \frac{(n-1)Q}{1+(n-1)Q}
\end{equation}
This gives
\begin{eqnarray}\label{firstorder2}
\langle 
W_R\rangle_{0}&\geq&\frac {n(n-1)}{\nu(R)}
\cdot\frac{Q}{1+(n-1)Q},\\
\langle 
W_R\rangle_{0}&\leq&
\frac {n(n-1)}{\nu(R)}
\cdot Q\ .
\end{eqnarray}

{} From Temple's inequality (see \cite{LY1998}, \cite{LSY1999})  we 
obtain the estimate
\begin{equation}\label{temple}E_{0}(n,\ell)
\geq (1-\varepsilon)\langle 
W_R\rangle_{0}\left(1-\frac{\mu \big(\langle
W_R^2\rangle_0-\langle W_R\rangle_0^2\big)}{\langle
W_R\rangle_0\big(E_{1}^{(0)}-\mu \langle W_R\rangle_0\big)}\right)
\end{equation}
where
\begin{equation}
E_{1}^{(0)}=\frac{\varepsilon\mu}{\ell^{2}}
\end{equation}
is the energy of the lowest excited state of $\varepsilon T^{{(n)}}$. 
This estimate is valid for $E_{1}^{(0)}/\mu > \langle 
W_R\rangle_0$, i.e., it is important that $\ell$ is not too big. 

Putting \eqref{firstorder2} and \eqref{temple} together we obtain 
the estimate
\begin{equation}\label{alltogether}E_{0}(n,\ell)\geq 
\frac{n(n-1)}{\ell^{2}}\,\frac {A(R)}{\nu(R)}\,
K(n)
\end{equation}
with
\begin{equation}\label{k}
K(n)=(1-\varepsilon)\cdot 
\frac{(1-\mfr{2R}/{\ell})^{2}}{1+(n-1)Q}\cdot
\left(1-\frac n{(\varepsilon\,\nu(R)/\ell^{2})-n(n-1)\,Q}\right)
\end{equation}
Note that $Q$ depends on $\ell$ and $R$, and $K$ depends on 
$\ell$, $R$ and $\varepsilon$ besides $n$.  We have here dropped the 
term  $\langle W_R\rangle_0^2$ in the numerator in \eqref{temple},  
which is  appropriate  for the purpose of a lower bound. 

We note that $K$ is monotonically decreasing in $n$, so for a given 
$n$ we may replace $K(n)$ by $K(p)$ provided $p\geq n$.  As 
explained in  \cite{LY1998} 
convexity of $n\mapsto n(n-1)$ together with 
superadditivity of $E_{0}(n,\ell)$ in $n$ 
leads, for $p=4\rho\ell^{2}$,  to an estimate for the energy of $N$ 
particles in the 
large box when the  side length $L$ is  an integer multiple of $\ell$: 
\begin{equation}\label{alltogether2}E_{0}(N,L)/N\geq \frac 
{\rho A(R)}{\nu(R)}\left (1-\frac 1{\rho\ell^{2}}\right) 
K(4\rho\ell^{2})
\end{equation}
with $\rho=N/L^3$.

Let us now look at the conditions on the parameters $\varepsilon$, $R$ 
and $\ell$ that have to be met in order to 
obtain a lower bound with the same leading term as the upper bound 
\eqref{upperbound}.

{}From \eqref{nu} we have 
\begin{equation}
\frac{A(R)}{\nu(R)}=\frac{4\pi}
{\left(\ln(R^{2}/a^{2})-1\right)}\left(1-{\rm 
O}((R_{0}^{2}/R^{2})\ln(R/R_{0})\right)
\end{equation}
We thus see that as long as $a<R<\rho^{-1/2}$ the logarithmic factor
in the denominator in \eqref{alltogether2} has the right form for a lower 
bound. Moreover, for 
Temple's inequality the denominator in the second factor in \eqref{k} 
must be positive. With $n=4\rho\ell^2$ and 
	$\nu(R)\geq {\rm(const.)} R^2\ln(R^2/a^2)\ {\rm for}\ R\gg R_{0}$,
this condition amounts to
\begin{equation}\label{templecond}
{\rm (const.)}\varepsilon \ln(R^2/a^2) /\ell^{2}>\rho^{2}\ell^{4}.
\end{equation}
The relative error terms in \eqref{alltogether2} that have to be $\ll 1$ 
are
\begin{equation}\label{errors}
	\varepsilon,\quad \frac{1}{\rho\ell^{2}},\quad 
\frac{R}{\ell},\quad\rho R^2,\quad
\frac{\rho\ell^4}{\varepsilon R^2\ln(R^2/a^2)}.
\end{equation}
We now choose
\begin{equation}
\varepsilon\sim|\ln(\rho a^2)|^{-1/5},
\quad \ell\sim \rho^{-1/2}|\ln(\rho a^2)|^{1/10},
\quad R\sim \rho^{-1/2}|\ln(\rho a^2)|^{-1/10}
\end{equation}

Condition \eqref{templecond} is satisfied since the left side is $>{\rm
(const.)}|\ln(\rho a^2)|^{3/5}$ and the right side is $\sim |\ln(\rho
a^2)|^{2/5}$. The first three error terms in \eqref{errors} are all of
the same order, $|\ln(\rho a^2)|^{-1/5}$, the last is $\sim
|\ln(\rho a^2)|^{-1/5}(\ln|\ln(\rho a^2)|)^{-1}$. With these 
choices, \eqref{alltogether2} thus leads to the following:

\begin{thm}[Lower bound]
For all $N$ and $L$ large enough such that $L>{\rm (const.)}
\rho^{-1/2}|\ln(\rho a^2)|^{1/10}$ and 
$N>{\rm (const.)}|\ln(\rho a^2)|^{1/5}$ with $\rho=N/L^2$, the 
ground state energy with Neumann boundary condition satisfies
\begin{equation} \label{lower}
E_{0}(N,L)/N\geq \frac{4\pi\mu\rho}{|\ln(\rho a^2)|}
\left(1-{\rm O}(|\ln(\rho a^2)|^{-1/5})\right).	
\end{equation}	
\end{thm}

In combination with the upper bound of Theorem 2.1 this also proves
\begin{thm}[Energy at low density in the thermodynamic limit]
\begin{equation}
\lim_{\rho a^2\to 0}\frac{e_0(\rho)}{4\pi\mu\rho|\ln(\rho a^2)|^{-1}}=1
\end{equation}
where $e_0(\rho)=\lim_{N\to\infty} E_0(N,\rho^{-1/2}N^{1/2})/N$.
This holds irrespective of boundary conditions.
\label{limitthm}
\end{thm}

{\it Remarks:} 1. It follows from the the remark at the end of Appendix A
that Theorem \ref{limitthm} is also valid for an infinite range potential
$v$ provided  that $v\geq 0$ and that  for some $R$ we have
$\int_{R}^{\infty} v(r)r\  dr <\infty$.

2. As in \cite{LY1998}, \cite{LSY1999} we could derive explicit bounds for
the error term in \eqref{lower}, but there is little reason to belabor
this point.

\appendix
\section{Appendix:  Definition and Properties of Scattering Length}

In this appendix we shall define and derive the scattering length and
some of its properties. The reader is referred to \cite{analysis},
especially chapters 9 and 11, for many of the concepts and facts we
shall use here. While we are interested in two dimensions, much of the
following is valid in all dimensions.

We start with a potential $\frac{1}{2}v(x)$ that depends only on the 
radius,
$r=|x|$, with $x\in \R^n$. For simplicity, we assume that 
$v$ has finite range; this condition can easily be relaxed, but we shall
not do so here, except for a remark at the end that shows how
to extend the concepts to infinite range, nonnegative potentials. 
Thus, we assume that
\begin{equation}
v(r) =0  ~~~~~~~~ {\rm for}~ r>R_0.
\end{equation}
We decompose $v$ into its positive
and negative parts,
$v=v_+ - v_-$, with $v_+,\ v_-\geq 0$,  and assume the following for
$v_-$ only (with $\epsilon >0$):
\begin{equation}
v_- \in \ \ \ \begin{cases} L^1(\R^1) & {\rm for}~ n=1 \\
L^{1+\epsilon}(\R^2)  & {\rm for}~ n=2 \\
L^{n/2}(\R^n) & {\rm for}~ n\geq3.
\end{cases}
\label{spaces}
\end{equation}
In fact, $v$ can even be a finite, spherically symmetric 
measure, e.g., a sum of
delta functions.

We also make the important assumption 
that $ \frac{1}{2}v(x)$ has no negative energy bound
states in $L^{2}(\R^n)$, which is to say we assume that for all $\phi
\in H^1(\R^n)$ (the space of $L^2$ functions with $L^2$ derivatives)
\begin{equation}
\int_{\R^n} \mu |\nabla \phi (x)|^2 + \frac{1}{2}v(x)|\phi(x)|^2~
d^nx \geq 0
\label{nobound}
\end{equation}

\begin{thm}\label{energy}
 Let $R>R_0$ and let $B_R \subset \R^n$ 
denote the ball $\{x: 0<|x| <R\}$ and $S_R$ the sphere $\{x: |x| =R\}$.
For $f\in H^1(B_R)$ we set 
\begin{equation}
\mathcal{E}_R[\phi] = \int_{B_R} \mu |\nabla \phi(x)|^2 + 
\frac{1}{2}v(x)|\phi(x)|^2
\label{E}
\end{equation}
Then, in the subclass of functions such that $\phi(x) =1$ for all $x\in
S_R$,
there is a unique function $\phi_0$ that minimizes $\mathcal{E}_R[\phi]$.
This function is nonnegative and spherically symmetric, i.e,
\begin{equation}\phi_0(x)=f_0(|x|)\end{equation}
with a nonnegative function $f_0$ on the interval $(0,R]$, and it satisfies
the equation
\begin{equation}\label{dist}
-\mu \Delta \phi_0(x) + \frac{1}{2}v(x)\phi_0(x) =0
\end{equation}
in the sense of distributions on $B_R$, with  boundary
condition $f_0(R)=1$.

For $R_0<r<R$
\begin{equation}
f_0(r) =  f_0^{\rm asymp}(r) \equiv 
\begin{cases} (r-a)/(R-a)  & {\rm for}\ n=1 \\
\ln (r/a)/\ln(R/a)  & {\rm for}\  n=2 \\
   (1-ar^{2-n})/( 1-aR^{2-n})& {\rm for}\  n\geq 3
\end{cases}
\label{asymp}
\end{equation}
for some number $a$ called the \underline{{\rm scattering length}}.

The minimum  value of $\mathcal{E}_R[\phi]$ is 
\begin{equation}
E= \begin{cases} 2\mu /(R-a) & {\rm for}\ n=1 \\
2 \pi \mu /\ln(R/a)  & {\rm for}\  n=2 \\
   2 \pi^{n/2} \mu a/[\Gamma(n/2)( 1-aR^{2-n})]& 
{\rm for}\  n\geq 3 .
\end{cases}
\label{emin}
\end{equation}
\label{minimizer}
\end{thm}

{\it Remarks:} 
1. Given that the minimizer is spherically symmetric for every $R$, it
is then easy to see that the $R$ dependence is trivial.  There is really
one function, $F_0$, defined on all of the positive half axis, such that 
$f_0(r) =F_0(r)/F_0(R)$. That is why we did not bother to indicate the 
explicit
dependence of $f_0$ on $R$. The reason is a simple one: If
$\widetilde{R} >R$, take the minimizer $\widetilde{f}_0$ for
$\widetilde{R} $ and replace its values for $r<R$ by
$f_0(r)\widetilde{f}_0(R)$, where $f_0$ is the minimizer for the $B_R$
problem.  This substitution cannot increase 
$\mathcal{E}_{\widetilde R}$.  Thus, by
uniqueness, we must have that $\widetilde{f}_0(r) =
f_0(r)\widetilde{f}_0(R)$ for $r\leq R$. \\

2. {}From \eqref{asymp} we then see that $f_0^{\rm asymp}(r)\geq 0$ for all
$r>R_0$, which implies that $a\leq R_0$ for $n\leq 3$ and 
$a\leq R_0^{n-2}$ for $n>3$.  \\

3. According to our definition \eqref{asymp}, $a$ has the dimension of a
length only when $n\leq 3$. \\

4. The variational principle \eqref{E}, \eqref{emin} allows us
to  discuss the connection between the
scattering length and $\int v$. We 
recall Bogolubov's perturbation
theory \cite{BO}, which says that to leading order in the density $\rho$, 
the
energy per particle of a Bose gas is $e_0(\rho)\sim 2\pi \rho \int v$,
whereas the correct formula in two-dimensions 
is $4 \pi \mu \rho |\ln(\rho a^2)|^{-1}$.
The Bogolubov formula is an upper bound (for all $\rho$) since it is the
expectation value of $H^{(N)}$ in the non-interacting ground state 
$\Psi \equiv 1$. Thus, we must have 
$\frac{1}{2} \int v \geq   |\ln(\rho a^2)|^{-1}$ when
$\rho a^2 \ll 1$, which suggests that 
\begin{equation}
\int_{\R^2} v \geq \frac{4\pi \mu }{\ln (R_0/a)}.
\label{2scattineq}
\end{equation}

Indeed, the truth of (\ref{2scattineq}) can be verified by using the 
function
$\phi(x) \equiv 1$ as a trial function in (\ref{E}). Then, using
(\ref{emin}),
$\frac{1}{2} \int v \geq E = 2\pi \mu/ \ln(R/a)$
for all $R \geq R_0$, which proves (\ref{2scattineq}). 
As $a \to 0$, (\ref{2scattineq}) becomes an equality, however,
in the sense that $(\int_{\R^2} v) \ln(R_0/a) \to 4\pi \mu$.

In the same way, we can derive the inequality of 
Spruch and Rosenberg \cite{SR} for dimension 3 or more:
\begin{equation}\label{3scattineq}
\int_{\R^n} v \geq \frac{4\pi^{n/2} \mu a}{\Gamma(n/2)}.
\end{equation}
(Here, we
take the limit $R\to \infty$ in \eqref{emin}).

In one-dimension we obtain (with $R=R_0$)
\begin{equation}\label{1scattineq}
\int_{\R} v \geq \frac{4\mu}{R_0 - a }.
\end{equation}

{\it Proof of Theorem \ref{energy}:} Given any $\phi\in H^1$ we can
replace it by the square root of the spherical average of $|\phi|^2$. This
preserves the boundary condition at $|x|=R$, while the   $v$ term in
\eqref{E} is unchanged. It also lowers the gradient term in (\ref{E})
because the map $\rho \mapsto \int (\nabla \sqrt \rho)^2$ is convex
\cite{analysis}.  Indeed,  there is a strict decrease unless $\phi$ is 
already
spherically symmetric and nonnegative.

Thus, without loss of generality, we may consider only nonnegative,
spherically symmetric functions. We may also assume that in the annular
region $\mathcal{A} = \{x \ : \ R_0\leq |x|\leq R\}$ there is some $a$
such that (\ref{asymp}) is true because these are the only spherically
symmetric, harmonic functions in $\mathcal{A}$.  If we substitute for
$\phi$ the harmonic function in $\mathcal{A}$ that agrees with $\phi$ at
$|x|=R_0$ and $|x|=1$ we will lower $\mathcal{E}_R$ unless $\phi$ is 
already
harmonic in $\mathcal{A}$.  (We allow the possibility $a=0$ for $n\leq
2$, meaning that $\phi =$ constant.)

Next, we note that $\mathcal{E}_R[\phi]$ is bounded below. If it were not
bounded then (with $R$ fixed) we could find a sequence $\phi^j$ such that
$\mathcal{E}_R(\phi^j) \to -\infty$.  However, if $h$ is a smooth
function on ${\mathbb R}_+$ with $h(r) =1 $ for $r<R+1$ and
$h(r) =0$ for $r>2R+1$ then the function $\widehat{\phi^j}(x) = \phi^j(x) $
for $|x|\leq R$  and $\widehat{\phi^j}(x) = h (|x|) $ for $|x|>R$ is a 
legitimate
variational function for the $L^2(\R^n)$ problem in (\ref{nobound}).  It
is easy to see that $\mathcal{E}_R[\widehat{\phi^j}] \leq
\mathcal{E}_R[\phi^j] +{\rm (const)} R^{n-2}$, and this contradicts
(\ref{nobound}) (recall that $R$ is fixed).

Now we take a minimizing sequence $\phi^j$ for $\mathcal{E}_R$ and
corresponding $\widehat{\phi^j}$ as above. By the assumptions on $v_-$ we
can see that the kinetic energy $T^j = \int |\nabla \phi^j|^2$ and $\int
|\phi^j|^2$ are  bounded. We can then find a subsequence of the 
$\widehat{\phi}^j$
that converges weakly in $H^1$ to some spherically symmetric
$\widehat{\phi}_0(x)=\widehat f_0(|x|)$. Correspondingly, $\phi^j(x)$ 
converges weakly in $H^1(B_R)$
to $\phi_0(x)=f_0(|x|)$.  The important point is that the term -$\int 
v_-|\phi^j|^2$ is
weakly continuous while the term $\int v_+|\phi^j|^2$ is weakly lower
continuous \cite{analysis}. We also note that $f_0(R) =1$ since the
functions $\widehat{\phi^j}$ are identically equal to $1$ for $R<|x|<R+1$ 
and 
the limit $\widehat{\phi}_0$ is continuous away from the origin 
since it is spherically symmetric and in $H^1$.

Thus, the limit function $\phi_0$ is a minimizer for  $\mathcal{E}[\phi]$
under the condition $\phi=1$ on $S_R$. Since it is a minimizer, it 
must be harmonic in $\mathcal{A}$, so $\eqref{asymp}$ is true.
Eq. \eqref{dist} is standard and is obtained by replacing 
$\phi_0$ by $\phi_0+\delta \psi$, where $\psi$ is any infinitely 
differentiable
function that is zero for $|x|\geq R$. The first variation in $\delta$
gives \eqref{dist}. 

Eq. \eqref{emin} is obtained by using integration by parts to compute
$\mathcal{E}_R[\phi_0]$.

The uniqueness of the minimizer can be proved in two ways. One way is to
note that if $\phi_0\neq \psi_0$ are two minimizers then, by the convexity
noted above, $\mathcal{E}_R [\sqrt{\phi_0^2 +\psi_0^2}) < 
\mathcal{E}_R[\phi_0]
+\mathcal{E}_R(\psi_0)$. The second way is to notice that all minimizers
satisfy \eqref{dist}, which is a linear, ordinary differential equation
 for $f_{0}$  on $(0,R)$ since all minimizers are spherically symmetric, as we noted.
But the solution of such equations, given the value at the end points, is
unique. \hfill   $\Box $

We thus see that if the Schr\"odinger operator on $\R^n$ with potential
$\frac{1}{2}v(x)$ has no negative energy bound state then  the
scattering length in (\ref{asymp}) is well defined by a variational
principle. Our next task is to find some properties of the minimizer
$\phi_0$. For this purpose we shall henceforth assume that $v$ is
\underline{\it nonnegative}, which guarantees (\ref{nobound}), of course.

\begin{lem} \label{lem1.1}
 If $v$ is nonnegative  then for all $0<r\leq R$ the
minimizer $\phi_0(x)=f_0(|x|)$ satisfies

A) 
\begin{equation} 
f_0(r) \geq f_0^{\rm asymp}(r),
\label{bound}
\end{equation}
where $ f_0^{\rm asymp}$ is given in (\ref{asymp})

B) 
$f_0(r)$ is a monotonically nondecreasing function of $r$.

C) If $v(r)\geq \widetilde{v}(r) \geq 0 $ for all $r$ then 
the corresponding
minimizers satisfy $f_0(r) \leq  \widetilde{f}_0(r)$ for all $r<R$.
Hence, $a >  \widetilde{a} \geq 0$.

\end{lem}

\begin{proof}
Let us define $f_0^{\rm asymp}(r)$ for {\it all} $0<r<\infty$ by 
\eqref{asymp}, and let us extend $f_0(r)$ to all $0<r<\infty$ by 
setting $f_0(r) = f_0^{\rm asymp}(r)$ when $r\geq R$.

To prove A) Note that $-\Delta \phi_0 = -\frac{1}{2}v\phi_0$, which implies
that $\phi_0$ is subharmonic (we use $v\geq 0$ and $\phi_0 \geq 0$, by 
Theorem
\ref{energy}). Set $h_{\varepsilon}(r)= f_0(r) - (1+\varepsilon) f_0^{\rm
asymp}(r)$ with $\varepsilon >0$ and small.  Obviously, $x\mapsto 
h_\epsilon(|x|)$ is
subharmonic on the open set $\{x: 0<|x|<\infty \}$ because $f_0^{\rm
asymp}(|x|)$ is harmonic there. Clearly, $h_{\varepsilon} \to -\infty$ as
$r\to \infty$ and $h_{\varepsilon}(R) = -\varepsilon$.  Suppose that
(\ref{bound}) is false at some radius $\rho <R$ and that $h_0(\rho) =
-c<0$. In the annulus $\rho < r < \infty$, $h_{\varepsilon}(r) $ has its
maximum on the boundary, i.e., either at $\rho $ or at $\infty$ (since
 $h(|x|)$  is subharmonic in $x$ ).  By choosing $\varepsilon$ sufficiently small and
positive we can have that $h_{\varepsilon}(\rho) < -2\varepsilon $ 
 and this contradicts the fact that the maximum (which is at least
$-{\varepsilon}$ ) is on the boundary.  

B) is proved by noting (by subharmonicity again) that the maximum 
of $f_0$ in $(0,r)$ occurs on the boundary, i.e., $f_0(r) \geq
f_0(r')$ for any $r'<r$. 

C) is proved by studying the function $g=f_0- \widetilde{f}_0$. Since
$f_0$ and $\widetilde{f}_0$ are continuous, the falsity of  C) implies
the existence some open subset, $\Omega \subset B_R$ on which $g(|x|) >0$.
On $\Omega$ we have that $g(|x|)$ is subharmonic (because $vf_0 >
\widetilde{v} \widetilde{f}_0$). Hence, its maximum occurs on the
boundary, but $g=0$ there. This  contradicts  $g(|x|)>0$ on $\Omega$.

\end{proof} 

{\it Remark about infinite range potentials:} If $v(r)$ is infinite
range and nonnegative it is easy to extend the definition of the
scattering length under the assumptions: 

1) $v(r) \geq 0$ for all $r$ and 

2) For some $R_1$ we have $\int_{R_1}^{\infty} v(r)r^{n-1}\ dr
<\infty$.  

If we cut off the potential at some point $R_0 >R_1$ (i.e., set
$v(r)=0$ for $r>R_0$) then the scattering length is well defined but it
will depend on $R_0$, of course. Denote it by $a(R_0)$. By part C of
Lemma (\ref{lem1.1}), $a(R_0) $ is an increasing function of $R_0$. 
However, the bounds (\ref{2scattineq}) and (\ref{3scattineq}) 
and assumption 2) above
guarantee that $a(R_0)$ is bounded above. 
(More precisely, we need  a simple modification of
(\ref{2scattineq}) and (\ref{3scattineq}) to the potential
$\widehat{v}(r) \equiv \infty $ for $r \leq R_1 $ and $\widehat{v}(r)
\equiv v(r)$ for $r> R_1$. This is accomplished by replacing the `trial
function' $f(x)=1$ by a smooth radial function that equals $0$ for $r<
R_1$ and equals $1$ for $r>R_2$ for some $R_2>R_1$.)  Thus, $a$
is well defined by 
\begin{equation} 
a= \lim_{R_0 \to \infty}a(R_0) \ . 
\end{equation}

\section{Appendix:  Proof of Dyson's Lemma \ref{dysonl} in Two 
Dimensions}

\begin{proof}
In polar coordinates, $r,\theta$, one has 
$|\nabla \phi|^2 \geq |\partial \phi /\partial r|^2$. Therefore, it 
suffices to 
prove that for each angle $\theta \in [0,2\pi)$, and with
$\phi (r,\theta)$ denoted simply by $f(r)$,
\begin{equation} \label{radial}
\int_0^{R(\theta)} \mu |\partial f(r) /\partial r|^2 + 
\frac{1}{2}v(r)|f(r)|^2 ~rdr \geq  
 \mu  \int_0^{R(\theta)}  U(r)|f(r)|^2 ~rdr,
\end{equation}  
where $R(\theta)$ denotes the distance of the origin to the boundary
of $\mathcal{B}$ along the ray $\theta$. 

If $R(\theta) \leq R_0$ then \eqref{radial} is trivial because the 
right side is zero while the left side is evidently nonnegative. (Here,
$v\geq0$ is used.)

If $R(\theta) > R_0$ for some given
value of $\theta$, consider the disc $\mathcal{D}(\theta)=
\{x\in \mathbb{R}^2 \ :\ 0\leq |x|\leq R(\theta) \}$ centered at the
origin in $\mathbb{R}^2$ and of radius $R(\theta) $.
Our function $f$ defines a spherically
symmetric function, $x\mapsto f(\vert x\vert)$ on
$\mathcal{D}(\theta)$, and \eqref{radial} is 
equivalent to 
\begin{equation}\label{disc}
\int_{{\cal D}(\theta)} \mu|\nabla f (|x|)|^2 + \frac{1}{2}v(r)|f(|x|)|^2
\end{equation}

Now choose some $R\in (R_0,\ R(\theta))$ and note that the left side of
\eqref{disc} is not smaller than the same quantity with 
${\cal D}(\theta)$ replaced by the smaller disc ${\cal D}_R=
\{x\in \mathbb{R}^2 \ :\ 0\leq |x|\leq R \}$. (Again, $v\geq 0$ is used.)
According to Appendix A, Theorem \ref{minimizer}, eq. \eqref{emin}, and
linearity in $|f|^2$,
this integral over  ${\mathcal D}_R$ is at least
$E(R)|f(R)|^2$. Hence, for every $R_0<R<R(\theta)$,
\begin{equation} \label{pointwise}
2\pi \int_0^{R(\theta)} \mu |\partial f(r) /\partial r|^2 + 
\frac{1}{2}v(r)|f(r)|^2 ~rdr \geq    E(R)|f(R)|^2 .
\end{equation}
The proof is completed by noting that $E(R)=   2\pi \mu /\ln(R/a)$,
by multiplying  both sides of \eqref{pointwise} by $U(R)R\ln(R/a)$ and, 
finally, integrating with respect to $R$ from $R_0$ to $R(\theta)$.
\end{proof}

\end{document}